\documentclass[12pt]{iopart}
\usepackage{float}
\usepackage{graphicx}
\usepackage[UKenglish]{babel}
\usepackage[colorinlistoftodos,prependcaption,textsize=tiny]{todonotes}
\usepackage{url}

\begin{document}
	\title{A novel method of synchronising two atomic clocks }
	\author{A. Walton$^{1,2}$, A. McGlone$^1$ and B. Varcoe$^1$}
	\address{$^1$ School of Physics and Astronomy, University of Leeds, Leeds, LS2 9JT, United Kingdom}
	\address{$^2$ Email: pyaw@leeds.ac.uk}
	\begin{abstract}
		In this paper we present a novel method for atomic clock synchronisation which combines digital error correction and phase tracking.   
		Microwave broadcasts are used to measure the difference in frequency between a pair of atomic clocks. We measure the phase errors in QPSK broadcasts that arise from desynchronisation of atomic clocks to a high level of precision. The rate of change of this phase error is used to find the relative error in frequency of the reference signals provided by the two clocks. We obtain an uncertainty of this rate of change in the region of $10^{-15}$ Hz for a measurement time of 10 minutes. The base stability measured using this technique is found to be $2.66 \times10^{-17}$ Hz, which makes it appropriate for applications such as measuring geodetic height to the centimeter level of accuracy.  
	\end{abstract}
	\maketitle
	
\section{Introduction}
Atomic clock synchronisation is a key component in a number of emerging areas of physics including the square kilometer array \cite{Wang2015}, atomic clock geodesy \cite{Gozzard2022,Beloy2021,Hoang2021,Makinen2021,McGrew2018,Mehlstaubler2018,Droste2013,Predehl2012,Bothwell2022,Zheng2022} and searching for physics beyond the standard model \cite{Barontini2021}.
The concept of tracking frequency difference between two atomic clocks over a long distance is therefore a well established concept.
There is, however, a difference between time synchronisation and frequency stability. Atomic clocks are precision oscillators that prioritise frequency stability over long length scales. In order to extract exact time measurements from oscillators, an addition system is required.  
For example, in navigation a global time reference can be maintained by an atomic clock, but propagation time from a set of reference points is used to calculate location. 
With the aim of atomic clock synchronisation, time differences are detected by a change in phase. 
Hence timekeeping which uses direct frequency measurements and time synchronisation enabled by phase changes are both important.

The normal method of Atomic Clock frequency comparison involves detecting the beat frequency between two oscillators. 
For optical clocks, frequency combs can be used to reach a long term stability in the region of $10^{-20}$ \cite{Comb}.
However, this requires an optical fibre, or, at the very least a line of sight free space connection. 
For locations in which neither of these are available such as rural areas or between distant locations, other methods are required. 

Recently we have shown that thermal noise in a broadcast channel can be exploited to create locally random correlated noise that can be used for secret key exchange \cite{Newton2019,Newton2020,Ghesquiere2021}. 
In this paper we propose an alternative use for thermally correlated noise, namely a new type of time synchronisation protocol using digital broadcast signals - or Quadrature Phase-Shift Keying (QPSK). 
QPSK is a protocol commonly used in digital radio broadcasting, such as with WiFi and Bluetooth, in which digital information is encoded in the phase of a wave packet. Binary values are transmitted as the absolute phase and transitions.
By then encoding a sequence of values, both the sequence and the transitions between them can be used to error correct.
Differences in frequency between a pair of clocks used as time references produce a measurable phase change in a received signal over time. 
To measure this, we broadcast the QPSK signal to two parties, Alice and Bob, with Alice and the source sharing a reference signal.
A broadcast signal includes both the frequency of the source and the absolute phase. The broadcast signal can therefore be used for time and frequency synchronisation between separate atomic clocks.
This carries the added benefit that error correction can be used to stabilise the timing in the event of signal loss. 
This protocol follows a similar method presented in refs. \cite{Sooudi:15} and \cite{Sooudi:17} where Binary Phase-Shift Keying (BPSK) was used together with an optical system for frequency transfer to achieve an Allan Variance of $10^{-14}$.

To analyse the effectiveness of the current method, the relative frequency stability is determined by calculating the uncertainty in a frequency estimate of a microwave atomic clock, and comparing this uncertainty to the broadcast frequency. With this method, we have achieved a relative frequency stability in the region of $10^{-15}$ Hz over 10 minutes using a 70MHz broadcast. In Section \ref{Method}, the basic experimental setup is described, with the results analysed in Section \ref{Results}.

\section{Method}\label{Method}

\begin{figure}[H]
	\centering
	\includegraphics[width=0.3\textwidth]{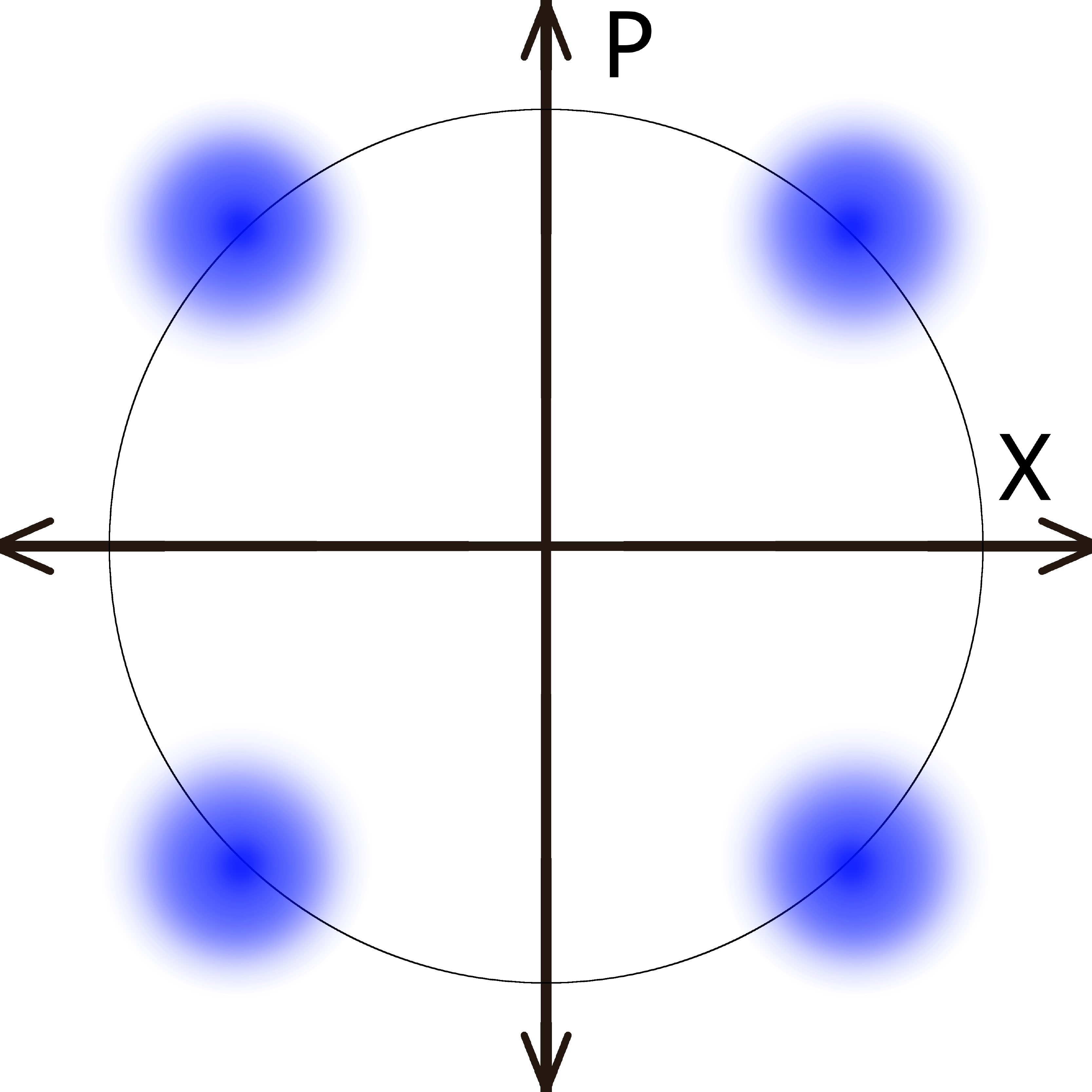}
	\caption{\textbf{QPSK.} A constellation diagram showing a QPSK broadcast. \label{fig:QPSK}}
\end{figure}
/todo{what do the axes in this graph represent? caption is a bit sparse, how do the axes relate to eqn 1}
In QPSK, the carrier wave is modulated in order to broadcast information. For a carrier wave given by:

\begin{equation}
 \Psi\left(t\right) = I\cos\left(\omega t\right) + Q\sin\left(\omega t\right),
\end{equation}
adjusting I and Q between two values gives four possible combinations, as shown in Figure \ref{fig:QPSK}, with thermal noise leading to clusters instead of single points on the constellation diagram. Data is encoded by assigning one of four binary values to each of the possible clusters.

\begin{figure}[H]
	\centering
	\includegraphics[width=0.5\textwidth]{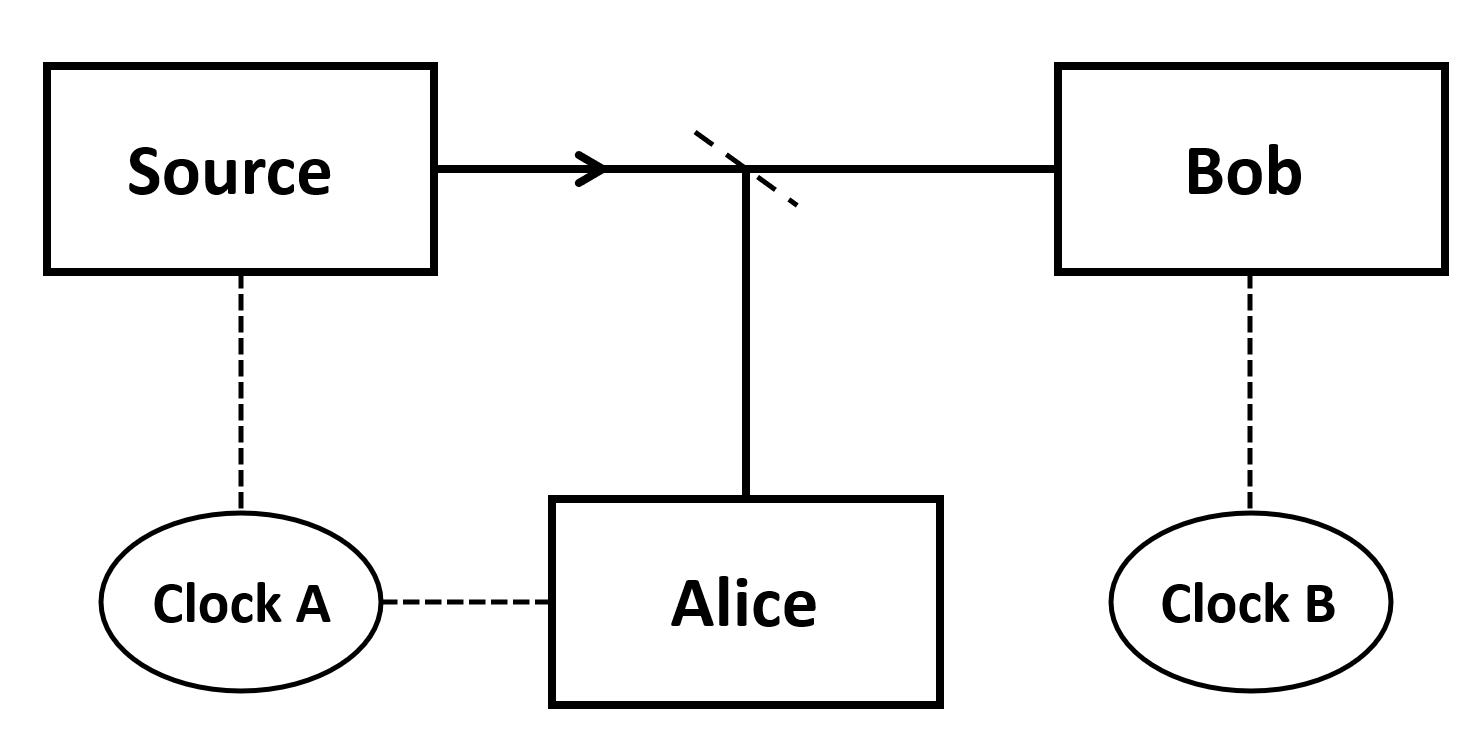}
	\includegraphics[width=0.5\textwidth]{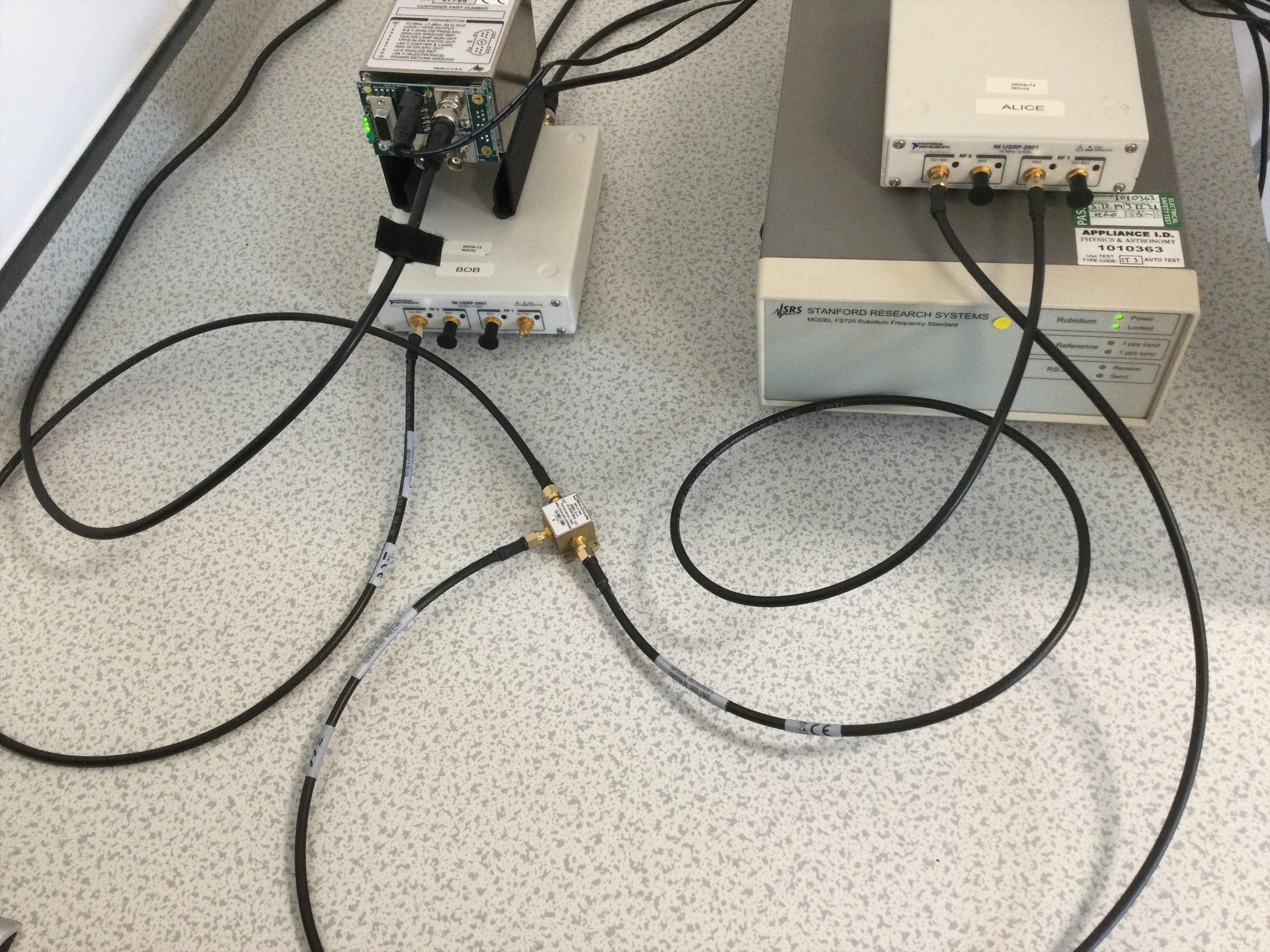}
	\caption{\textbf{Time Distribution.} The protocol used to compare frequency from two clocks. Alice and the source share a primary clock, while Bob uses a secondary clock, with the source broadcasting a QPSK signal. The photo shows the USRP for Alice and the source, with their shared clock (right), and Bob's USRP with it's connected clock (left). \label{fig:Protocol}}

\end{figure}

If a QPSK signal is broadcast to two parties, Alice and Bob, and perfectly synchronised clocks (or a single shared clock) are used to track time, we expect measurements for both parties to match the form shown in the constellation diagram in Figure \ref{fig:QPSK}. We implement this by using a USRP-2901 to broadcast the initial signal, with a power splitter directing the signal to Alice, who uses the same USRP as the source, and Bob, who measures with a separate USRP. This setup is displayed in Figure \ref{fig:Protocol}.

With this setup, Alice and the source share a reference 10MHz signal from a FS725 rubidium frequency standard, while Bob uses a PRS10C clock. Commonly, in this type of system, synchronisation is ensured through the use of signal processing tools such as a phase-locked loop. If this is not present, Bob's clusters will rotate about the origin in phase space as differences in frequency manifest as a change in phase over time, which is an undesirable effect when attempting to encode data in the phase of a broadcast as is done is phase-shift keying.

When broadcasting a QPSK signal without synchronised clocks, the rate of cluster rotation depends on the magnitude of the frequency error, where an error of 1Hz in the carrier frequency corresponds to a 1Hz rotation of Bob's clusters about the origin. By tracking Bob's average phase measurements over time, the rate of rotation is calculated, which gives the error in frequency between the pair of clocks.

\begin{figure}[H]
	\centering
	\includegraphics[width=0.5\textwidth]{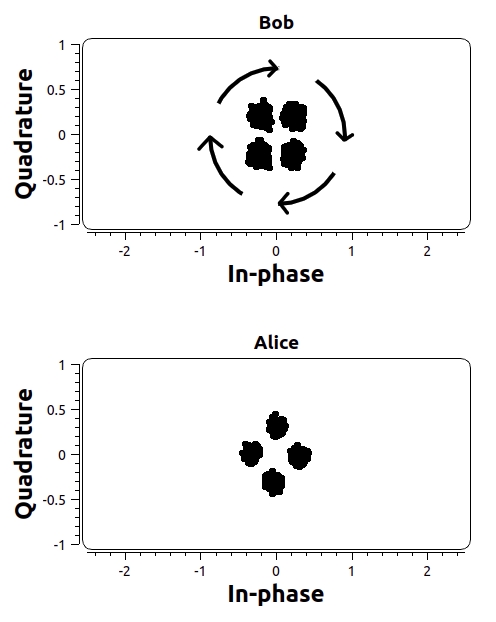}
	\caption{\textbf{GNURadio readings.} A snapshot of a QPSK broadcast performed by GNURadio at 70MHz. The readings for Alice remain centered on four points as is typical for a QPSK broadcast, while Bob's clusters rotate about the origin at an approximately constant rate. \label{fig:Clusters}}
\end{figure}

We use GNURadio to broadcast a signal typical of a QPSK protocol to Alice and Bob as seen in Figure \ref{fig:Clusters}. Due to the difference in reference frequencies and the removal of signal processing blocks typically used to correct phase errors, a gradual rotation is observed in Bob's clusters as well as a static phase offset.

While Alice's data is not necessary for this analysis, measurements that fall into each of the clusters observed by Alice correspond to measurements in each of Bob's clusters. As Alice's clusters remain stationary due to sharing a frequency reference with the source, this gives a simple method of tracking which of Bob's measurements belong to which cluster over time, and therefore assists in measuring rotation frequency. While it is possible to track rotation speed purely through Bob's measurements, if the noise or rotation speed becomes larger, it becomes increasingly difficult to discern which measurements belong to which cluster as the cluster edges begin to overlap.

This initially presents a problem as Alice and Bob's strings of measurements are not naturally aligned, and adjustments are necessary in order to fix the time delay. To align Alice and Bob's data strings, we rely on Hanbury Brown and Twiss correlations \cite{Ragy_2013}, as has been done in a similar experiment concerning measurements from a shared thermal source \cite{Walton1}. We compare the amplitudes of each measurement in a sample performed by Alice and Bob, and offset the data strings until we find a point of high $(R > 0.8)$ correlation, at which time the data strings are aligned and the time delay is removed. This point of high correlation exists due to the Hanbury Brown and Twiss effect, and gives a simple indication of the size of the time delay between Alice and Bob.

\section{Results}\label{Results}

\subsection*{A Shared Clock}

We begin with an idealised setup using a single clock, splitting a 10 MHz signal from the FS725 such that Alice, Bob, and the source each share a reference. This should provide a limit on the precision of this method as performed with our current equipment. This sets Clock A = Clock B in Figure \ref{fig:Protocol}.

As we are only interested in the rate at which Bob's clusters are rotating, we first simplify Bob's measurements by choosing one of the received clusters, and applying a rotation of $\frac{\pi}{2}$, $-\frac{\pi}{2}$, or $\pi$ as necessary to map measurements from the other clusters to the chosen one. Bob records strings of in-phase and quadrature components of the received signal and calculates the phase angle of each measurement pair. The measurements are split into blocks of 5000, and we track the mean value of each block over time.

\begin{figure}[H]
	\centering
	\includegraphics[width=0.8\textwidth]{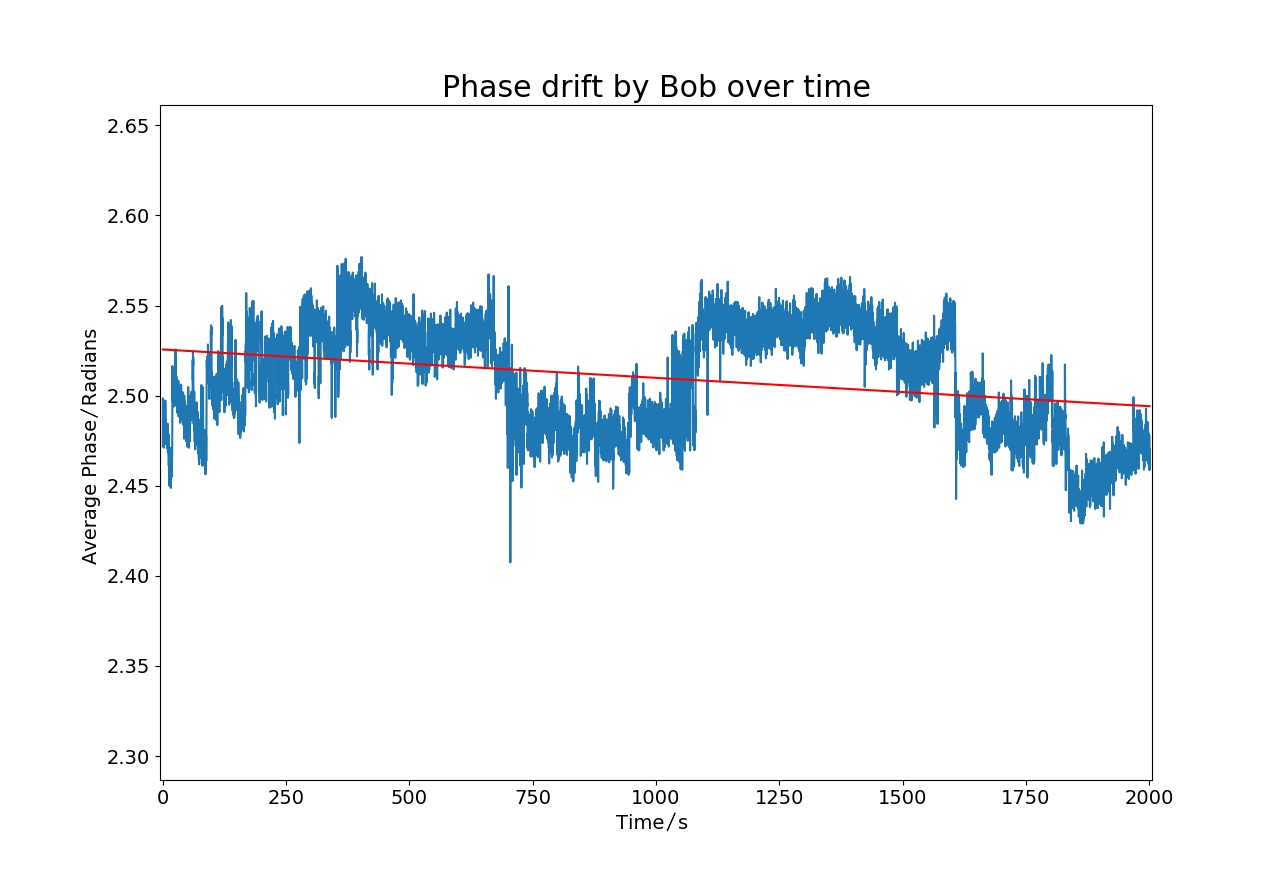}
	\caption{\textbf{Shared Clocks.} The results of the protocol when performed with a single clock at 3 GHz. We still see a gradual phase change over time, though the effect is very minor. \label{fig:SingleClock}}
\end{figure}

\begin{figure}[H]
	\centering
	\includegraphics[width=0.8\textwidth]{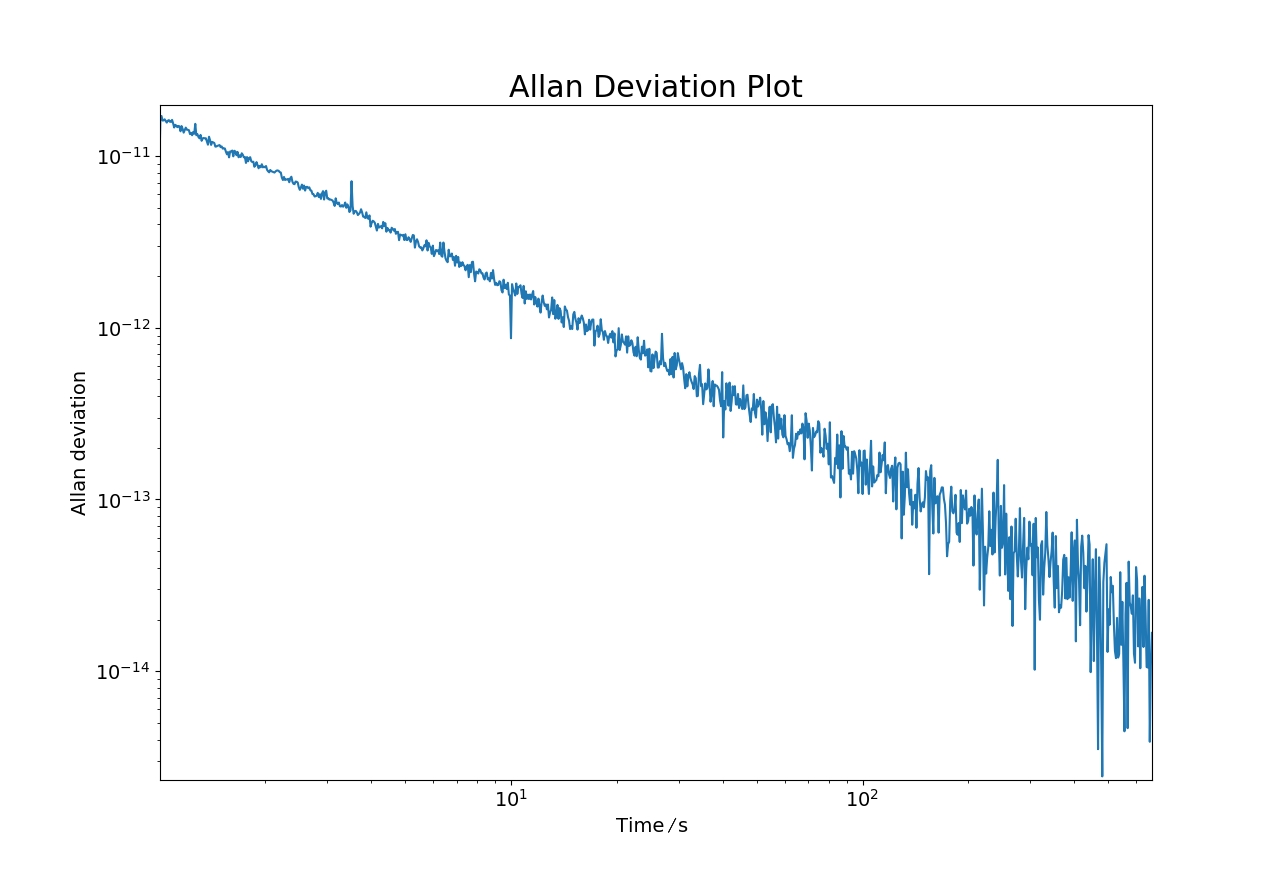}
	\caption{\textbf{Allan Deviation.} An Allan Deviation plot of the single clock phase measurements from Figure \ref{fig:SingleClock}, performed using a 3 GHz broadcast. \label{fig:Allan}}
\end{figure}

For this initial test, we broadcast a QPSK signal to Alice and Bob at a frequency of 3 GHz, and calculate the phase of each measurement, with the mean values of each block of 5000 measurements plotted in Figure \ref{fig:SingleClock}. The Allan deviation of the same raw data is given in Figure \ref{fig:Allan}. We see that the average phase remains approximately stable throughout the measurement time, as is expected given the shared clock. After converting the drift in phase measurements into frequency, we find the uncertainty in gradient as a factor of the 3 GHz broadcast frequency is $2.66\times10^{-17}$ after 2000.5 seconds. This acts as a limit on uncertainty for this time frame with the equipment we have access to. Later tests involve multiple clocks over a smaller measurement period, so we do not expect to attain this level of precision in frequency comparison. A zoomed view of the first 30 seconds of data shows that there are small oscillations not clear in the full view. This is a Fourier component that arises from the mapping of the rotation onto a rectangular coordinate system. This oscillation will therefore occur at the same rate as the rate of change of slope and provides an additional fit component.

\subsection*{Clock Comparisons}

\begin{figure}[H]
	\centering
	\includegraphics[width=0.8\textwidth]{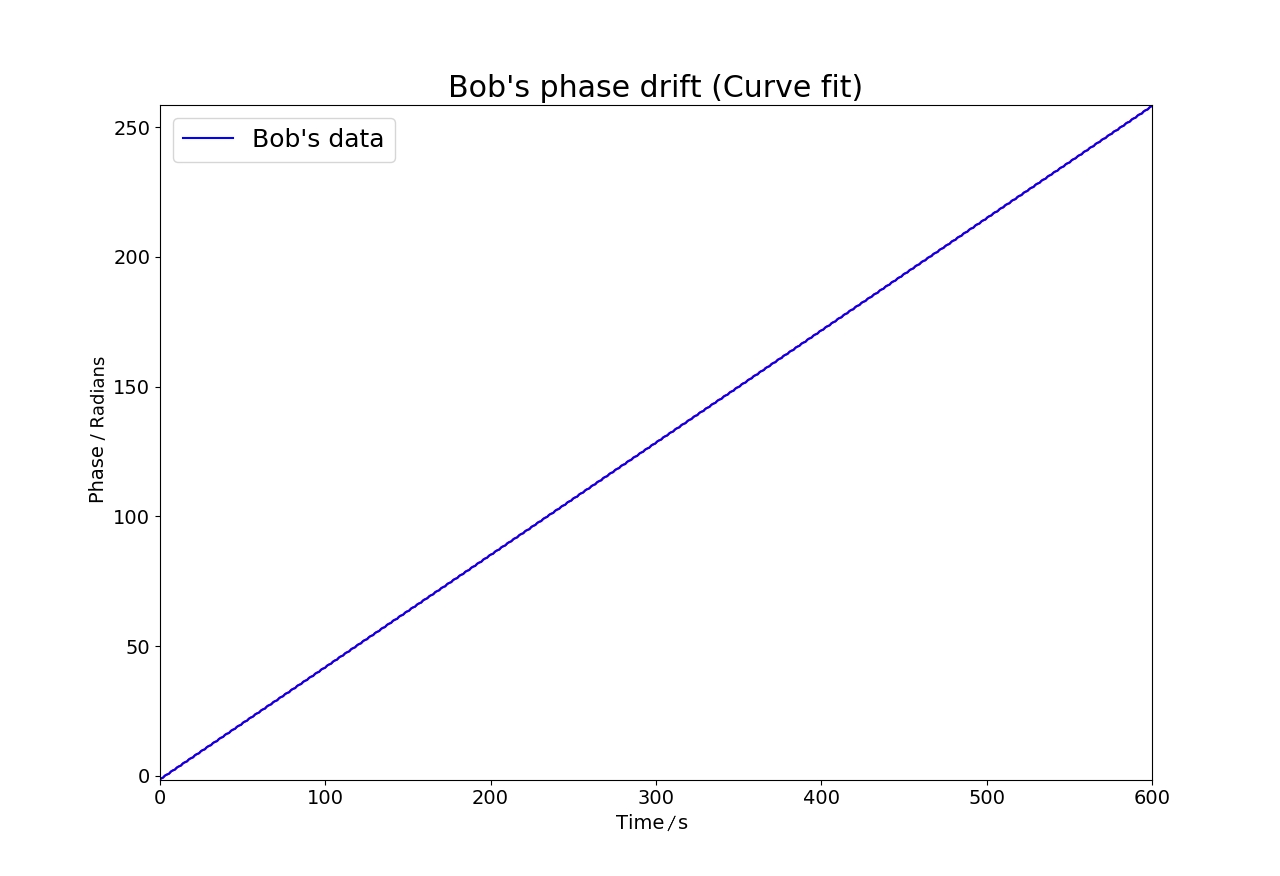}
	\includegraphics[width=0.8\textwidth]{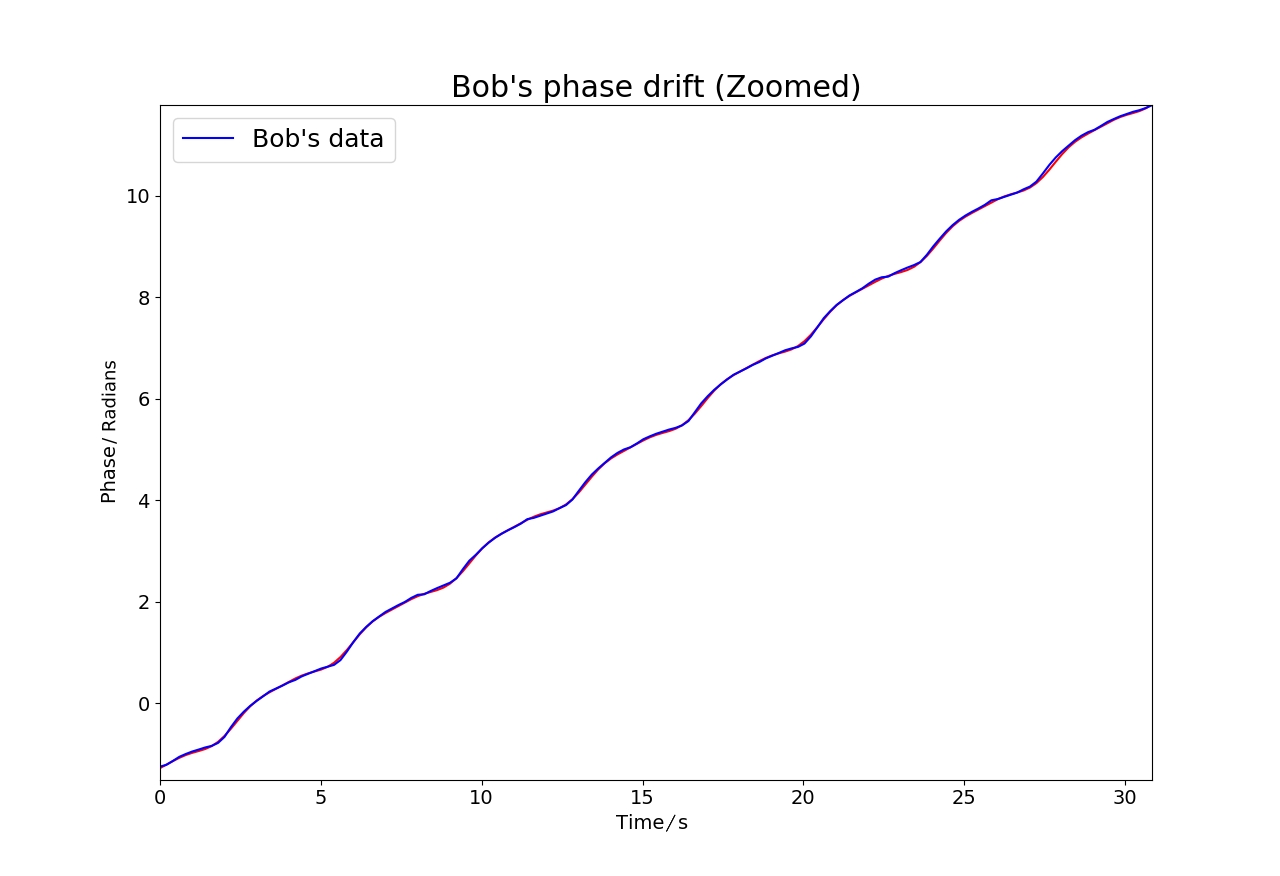}
	\caption{\textbf{Phase drift.} Bob's phase is observed to increase over time, the rate of change of which will be used to calculate the frequency error. A zoomed view of the first 30 seconds of data makes visible oscillations not clear in the full view. This is a Fourier component that arises from the mapping of the rotation onto a rectangular coordinate system. This oscillation will therefore occur at the same rate as the rate of change of slope and provides an additional fit component. \label{fig:BestFitCurve}}
\end{figure}

We now set up the equipment as displayed in Figure \ref{fig:Protocol}, with Alice and the source sharing a primary clock, while Bob relies on a secondary clock. For this test, we set the broadcast frequency to 70 MHz and record for 10 minutes (599.5s). The results from this are shown in Figure \ref{fig:BestFitCurve}, which includes a closer view to display smaller changes in measurements.

\begin{figure}[H]
	\centering
	\includegraphics[width=0.8\textwidth]{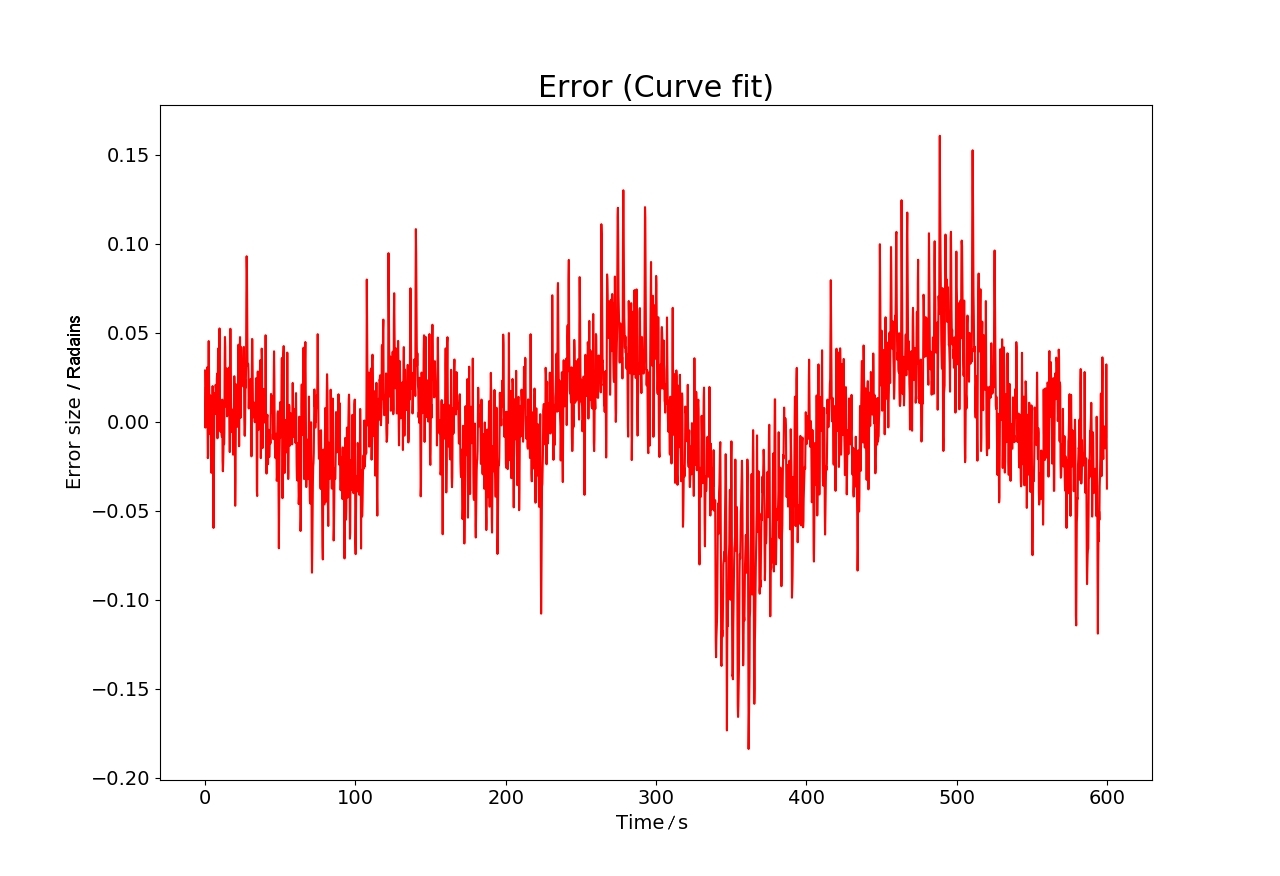}
	\caption{\textbf{Best fit error.} The residuals observed in the best fit line given in Figure \ref{fig:BestFitCurve}. \label{fig:BestFitError}}
\end{figure}

While we see the expected change in average phase over time due to the frequency difference, the rate of change is not linear, with periodic oscillations visible in the zoomed in version of the plot. Despite some adjustments to account for the majority of these oscillations, errors remain observable in the plot of the best fit curve residuals in Figure \ref{fig:BestFitError}. While this specific experiment showed regular oscillations on top of a constant gradient, this additional behavior proved difficult to characterise as the manner in which the data deviated from a straight line fit was not consistent as the frequency was changed. This is visible in the single clock variant of the experiment in Figure \ref{fig:SingleClock}, in which the data does not follow such a clear repeating pattern.

To achieve a good fit to the data the fit also requires that the Fourier components (a series in $\cos^2$ elements) are included in the the fit:

\begin{equation}
    \phi\left(t\right) = a~t + b + c\cos^2\left(2~a~t+\phi_1 \right) + f\cos^2\left(4~a~t+\phi_2\right)
\end{equation}

\begin{figure}[H]
	\centering
	\includegraphics[width=0.9\textwidth]{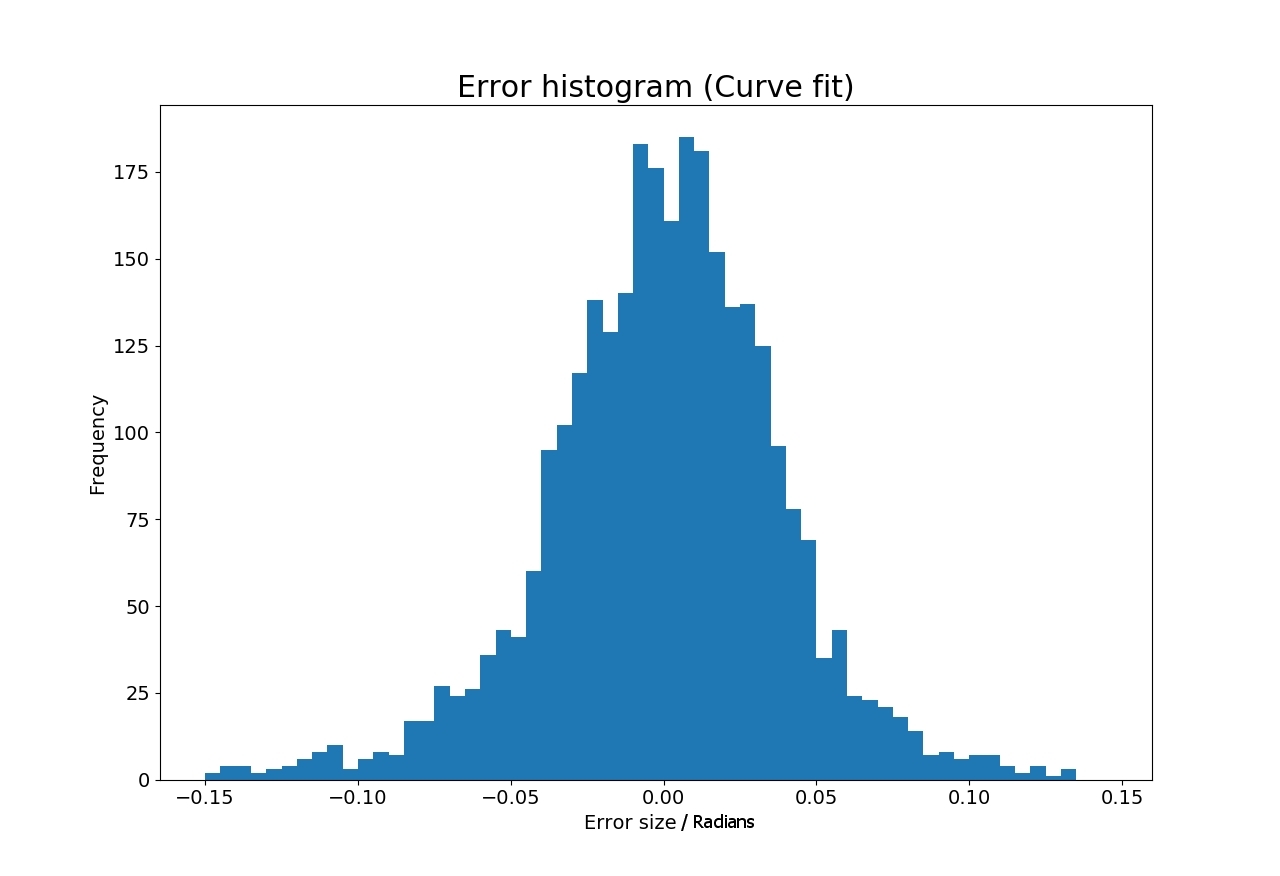}
	\caption{\textbf{Phase error.} A histogram compiling the errors observed in the best fit line given in Figure \ref{fig:BestFitError}. An approximate normal distribution suggests that the additional oscillations have been sufficiently accounted for, and the straight line component of the gradient can be analysed. \label{fig:Hist}}
\end{figure}

Residual errors are presented in Figures \ref{fig:BestFitError} and \ref{fig:Hist} in which we see that the residual errors are close to Gaussian and therefore the fit is good. The angular rate of change extracted by this fit is $0.4325$rad/s with an uncertainty of $4.2\times10^{-6}$, which is the angular velocity at which Bob's clusters rotate, and translates to a frequency error of $0.0688 \pm 6.669\times 10^{-7}$ for a 10 minute, 70MHz broadcast, which equates to a precision measurement of the 10MHz reference in the receiver clock frequency of $10,000,000.00983$Hz with an relative uncertainty of
\begin{equation}
    \frac{\delta f} {f_{10MHz}}=9.55\times10^{-15}.
\end{equation}

\section{Conclusions}\label{Conclusions}

We have developed a frequency exchange system that is capable of achieving a relative uncertainty in frequency of $9.55\times10^{-15}$ Hz over 600 seconds for a 70 MHz broadcast. 
The measurements are made via analysis of a typical QPSK broadcast with the phase-locked loop system removed and replaced with a reference clock. 
The accumulated difference in reference frequency results in a drift of phase over time, which can be measured and calculated. 
We find that for the equipment used in this example that the ultimate limit of precision is $2.66\times10^{-17}$ Hz for an experimental duration of 2000 seconds using a 3 GHz broadcast. Higher broadcast frequencies are not currently within range of the equipment used in this testing, and are a limiting factor on our attainable precision for a set time frame.

We also find that the rate of change of phase is not constant, instead displaying a periodic oscillation which becomes more chaotic as frequency is increased. This places a limit on the possible precision of this method due to the inherent instability of the clock. Ideally, future work would allow for a full characterisation of the rate of change of phase which is valid up to the 2-2.5GHz region, in which microwave communication equipment commonly operates.

Increasing the broadcast frequency increases the rotation speed of the clusters due to errors in reference frequencies being amplified, this allows for more precise measurements. Alternatively, more stable clocks will reduce the additional oscillations seen in the zoomed view in Figure \ref{fig:BestFitCurve}, allowing for a more consistent change of phase, and therefore a smaller uncertainty in the straight line gradient estimate. 
Given access to more accurate time keeping equipment, it therefore seems reasonable that this method could be used to make geodetic height measurements with an accuracy of 10cm and moving to an optical communications method operating at 100GHz could lower this even further to a sub-cm scale. 


\section*{Acknowledgements}
This work was supported by the Northern Triangle Initiative Connecting capability fund as well as funding from the UK Quantum Technology Hub for Quantum Communications Technologies EP/M013472. The data used to plot the graphs in Figures \ref{fig:SingleClock} and \ref{fig:BestFitCurve} is available from the Research Data Leeds Repository with the identifier \url{https://doi.org/10.5518/1193} \cite{data}.

\section*{References}
	
\bibliographystyle{unsrt}
\bibliography{Reference}
\end{document}